\documentclass[aps,prl,preprint,floatfix]{revtex4}
\usepackage{graphicx}
\usepackage{subfigure} 

\begin{document}

\title{Pairs of Gold Electrodes with Nanometer Separation Performed over SiO$_2$ Substrates with a Molecular Adhesion Monolayer } 

\author{Ajit K. Mahapatro$^1$, Subhasis Ghosh$^2$, and David B. Janes$^1$}
\affiliation{ 1. School  of Electrical and Computer Engineering, Purdue University, West Lafayette, Indiana 47907, USA  \\
2. School of Physical Sciences, Jawaharlal Nehru University, New Delhi 110067, India.\\
Contact Emails: janes@ecn.purdue.edu, ajit@ecn.purdue.edu}

\begin{abstract} 
Pairs of electrodes with nanometer separation (nano-gap) are achieved through an electromigration-induced break-junction (EIBJ) technique at room temperature. Lithographically defined gold (Au) wires are formed by e-beam evaporation over oxide coated silicon substrates silanized with (3-Mercaptopropyl)trimethoxysilane (MPTMS) and then subjected to electromigration at room temperature to create a nanometer scale gap between the two newly formed Au electrodes. The Si-O-Si covalent bond at the SiO$_2$ surface and the Au-sulphur (Au-S) bond at the top evaporated Au side, makes MPTMS as an efficient adhesive monolayer between SiO$_2$ and Au. Although the Au wires are initially 2$\mu$m wide, gaps with length $\sim$1nm and width $\sim$5nm are observed after breaking and imaging through a field effect scanning electron microscope (FESEM). This technique  eliminates the presence of any residual metal interlink in the adhesion layer (chromium or titanium for Au deposition over SiO$_2$) after breaking the gold wire and it is much easier to implement than the commonly used low temperature EIBJ technique which needs to be executed at 4.2 K. Metal-molecule-metal structures with symmetrical metal-molecule contacts at both ends of the molecule, are fabricated by forming a self-assembled monolayer of -dithiol molecules between the EIBJ created Au electrodes with nanometer separation. Electrical conduction through single molecules of 1,4-Benzenedimethanethiol (XYL) is tested using the Au/XYL/Au structure with chemisorbed gold-sulfur (Au-S) coupling at both contacts.  
\end{abstract}

\maketitle


% Will be ignored for other modes.

%\baselineskip=24pt

\section{Introduction}

Efficient device fabrication requires developing a suitable contact structure for characterization of few-molecule systems. Transport properties of single molecules are difficult to study by probing them between two symmetrical metal contacts due to the lack of a means to define the electrode structures with separations in molecular dimension.  Efforts have been made to study the electrical conductivity through molecules by fabricating metal-molecule-metal (M-M-M) structures.  Two general approaches involving pre-formed contacts have been employed.  The first involves vertical device structures (VDS), where a self-assembled monolayer (SAM) of molecules is prepared on a metallic surface and a second contact is made using scanning tunneling microscopy\cite{sd97,ege03} (STM), gold cluster assisted STM\cite{rpa96}, flip chip\cite{cmw04}, and cross wire technique\cite{jgk02}. The second approach involves lateral device structures (LDS), where molecules with -thiol functional groups are coupled to metal electrodes by placing them across a pair of electrodes with nanometer separation.  The electrodes are formed using mechanically controlled break junction\cite{mar97,ck99} (MCEB), electromigration-induced break junctions (EIBJ)\cite{wl02,jp02,lhy00,kib04} at low temperature (4.2K), and electrodeposition\cite{afm99,czl00,mmd03}. In VDSs the top contact to the molecule is made by a physical process, whereas the bottom contact is chemically coupled to the molecule. Here the presence of any asymmetry due to different contact structures at both ends of the molecule, can not be avoided in electrical transport measurements. In LDSs that have been demonstrated to create a nano-gap between two electrodes, the Au wires are either free standing microbridges\cite{ck99,jr02} or formed over SiO$_2$ substrates with metal (titanium\cite{kib04} or chromium\cite{hp99}) adhesion layer\cite{lhy00}. In these cases, after making the break junction the residual interlink of metallic adhesion layer, in devices employing a metallic adhesion layer over SiO$_2$, and the cantilever effect of the suspended Au electrodes, in the case of free standing structures, can not be ruled out. 

In this paper we present a simple method of fabricating a pair of electrodes with nanometer sized separation (nano-gap) between them, through an EIBJ technique performed at room temperature.  The starting structure consists of Au wires formed over thermally oxidised silicon substrates coated with a molecular adhesive monolayer of (3-Mercaptopropyl)trimethoxysilane (MPTMS). Electrical conduction through single molecule (1,4-Benzenedimethanethiol (XYL)) in a M-M-M structure (Au/XYL/Au) with nominally symmetric metal-molecule (Au-S) contacts is studied using Au nano-electrodes formed by this EIBJ process.

\section{Fabrication and Characterization of the nanometer sized electrode separation}

The 5000$\AA$ thermal oxide coated silicon(Si) substrates from Silicon Quest International, Inc., were silanized with MPTMS procured from Aldrich and Co., USA.  In silanization, the SiO$_2$ substrate was processed as follows: i) pirana clean, ii) oxygen plasma, iii) hydroxylation and iv) four hours of exposure to the MPTMS gaseous molecules inside a vacuum desiccator containing a open bottle of MPTMS. The silanised SiO$_2$ samples were transferred immediately into the evaporation chamber for Au layer deposition by e-beam evaporation at room temperature. The detailed experimental procedure and the chemical processes involved are explained elsewhere\cite{akm05}.  
The silane functional group of the MPTMS forms Si-O-Si covalent bonds with the SiO$_2$ surface silanols, leaving the -thiol group (-SH) of the molecule on the top, which on Au deposition forms Au-S bond, strong enough to hold the Au thin film tightly. 
A cartoon showing the Au film over a MPTMS capped SiO$_2$ substrate is shown in Fig.1(b). For electrical connections, each end of the lithographically defined Au wire is connected to a thick (4000$\AA$) Au pad layer deposited with e-beam evaporation (shown in fig.1(c)). In all depositions the pressure was maintained at $2-3 \times 10^{-7}$ Torr and the rate of deposition was controlled at 1$\AA$/sec, which was monitored using a quartz crystal based thickness monitor. The position of the nano-gaps are controlled by placing a notch structure at the middle of the Au wire(shown in fig.1(c)). Following electromigration, the nanogaps are characterized through field effect scanning electron microscope (FESEM), and current-voltage (I-V) characteristics. 

%\begin{figure}
%\centering
%\includegraphics[width=5in]{fig-1.eps}
%%{expt_fig.eps}
%%,height=1.1in
%\includegraphics[width=2.5in]{IEEE_nano_fig-1.eps}
%\caption{(a) Molecular formula of a MPTMS molecule. (b) A schematic diagram of MPTMS monolayer adhesion that forms Si-O-Si covalent bonds with the SiO$_2$ surface silanols and the Au-S bond with the evaporated Au thin layers. (c) Microscopic image of a 200$\AA$ thick lithographically defined Au wire with a notch at the middle and connected to 4000$\AA$ thick Au pad layers at both ends for electrical probing.}
%\label{fig_1}
%\end{figure}
%\begin{figure}
%\centering
%\includegraphics[width=5in]{fig-2a.eps}
%%{fig-2a2.eps}
%%{IEEE_nano_fig-2d.eps} 

%\vspace{0.4cm}

%\includegraphics[width=5in]{fig-2b.eps}
%{tun_hist.eps}
%\caption{(a) Current - voltage characteristics in a Au wire during EIBJ. Inset [A] shows the threshold current and voltage values at break point for some EIBJ devices, and [B] shows the I-V characteristics of an empty gap, solid line is the Fowler Northem fit to the experimental data (open circles). (b) Histogram of the conductances observed in 125 nano-gap EIBJ (Au/Empty-gap/Au) devices.}
%\label{fig_2}
%\end{figure}

A linearly increasing voltage (V) is ramped across the 200$\AA$ thick Au wires in steps of 20mV. At a threshold voltage $V_{th}$ (corresponding current density $J_{th}$), $\sim 10^7$ order decrease in current due to the formation of a gap is observed, where Ohmic current through the Au wires change to tunneling current through the nano-gap created between two newly formed Au electrodes. I-V characteristics through a Au wire during electromigration are shown in Fig.2(a). The resistances of the Au wires, which were 35-40 $\Omega$ before electromigration, change to $\sim G\Omega$ after the break. The empty gap current is in excellent agreement with the Fowler-Northeim tunneling expression\cite{smz2nd} and is shown in the inset of fig.1(a). 
The break occurs when the driving voltage exceeds 1.7V and a current over 50mA is flowing through the wire.
At this point, a gap is formed due to the physical motion of atoms (Au) out of the high-current density areas, due to electromigration. The movement direction is the net result of (i) the direct force due to the direct action of the external field on the charge of the migrating ion and (ii) the wind force due to the scattering of the conduction electrons by the metal atom under consideration. The threshold values of the current($J_{th}$) and voltage($V_{th}$) at the break point varies from wire to wire (plotted in fig.2(b)). This indicates that electromigration is a combined effect of the local Joule heating experienced at the notch and the net force between migrating ions and following electrons. Fig. 2(c) represents the histogram of the spectrum of conductances from $\sim \mu$S to $\sim$ pS, observed in some nano-gaps formed through EIBJ. 
Tunneling conductance of $\sim$nS is known to be due to approximately $\sim$ 1nm electrode separation\cite{czl00,afm99,jr02} and decreases exponentially as separation distance increases. The length of the gap is related to the tunneling current by the expression\cite{czl00} $I_{tun} \propto exp(-k d)$, where $I_{tun}$ is the tunneling current, ${d}$ is the gap length and ${k}$ is a constant. Hence from the tunneling conductance of the empty-gaps, the separation distance can be easily estimated.  
Although the Au wires are initially 2µm wide, imaging using the FESEM indicates final gaps(the regions with shortest separation distance between newly created Au electrodes) of length $\sim$1nm over $\sim$5nm width. Fig.3 shows the FESEM image for a pair of Au electrodes with 1nm separation observed in a EIBJ device. It is found that 20$\%$ of the total break junctions fabricated exhibit gaps of length $\sim$1-2nm of what appears to be silicon dioxide, 40$\%$ are $\sim$2-5 nm gap and the rest of the junctions show gaps between 5-15nm. The molecular adhesive monolayer eliminates the residual metal interlink between two newly created Au electrodes after breaking the Au wire, which is observed is structures using a layer of titanium as an adhesion layer for the Au wires. 

%\begin{figure}
%\centering
%\includegraphics[width=5in]{fig-3.eps}
%%height=2.5in]{brk_FESEM_1.eps}
%\caption{FESEM image of an EIBJ that shows a $\sim$1nm gap created between two Au electrodes. Although the Au wires are initially 2$\mu m$ wide (fig.a), gap of length $\sim$1nm and width $\sim$5nm is observed (fig.b) after the break junction. Image (b) is the zooming picture of the selected portion in image (a).}
%\label{fig_3}
%\end{figure}

%\section{Conduction through XYL molecule with symmetric Au-S contacts}

A dithiol molecule (XYL) is deposited between Au nanogap electrodes by immersing the Au electrical break junction chip into a 1mM solution of XYL in ethanol. In this case, the metal-molecule contacts are through chemisobed Au-S coupling, which forms a nominally symmetric contacts at both ends of the molecule. Au-S is known to be a strong bond and frequently used in molecular electronics\cite{mar97,jp02} for probing molecules with -SH end groups. Fig. 4 represents the symmetric I-V characteristics of Au/XYL/Au device at room temperature. An increase in current of $\sim 10^{5}$ is observed in these devices, compared to the currents through the empty-gap devices. This increase in conductivity is observed only in devices with gaps of $\sim$1-2 nm, but not in devices with gap larger than 2nm, as determined from empty-gap conductivity levels and subsequent FESEM imaging.

%\begin{figure}
%\centering
%\includegraphics[width=5in]{fig-4.eps}
%\caption{Room temperature I-V characteristics of Au/XYL/Au (squares) and Au/Empty-gap/Au (circles) structures.}
%\label{fig_6}
%\end{figure}

\section{Conclusions}

In conclusion, we present a simple method to achieve a pair of  electrodes with nanometer separation through EIBJ performed at room temperature.  The structure involves lithographically defined Au wires formed over SiO$_2$ substrates silanised with MPTMS.   The MPTMS acts as an efficient molecular adhesion monolayer between Au and SiO$_2$. This technique of fabricating M-M-M structures with symmetric metal-molecule contacts, eliminates the presence of any residual metal interlink in the adhesion layer (chromium or titanium) after breaking the gold wire and it is much easier to implement than the commonly used low temperature breaking technique, which needs to be executed at 4.2 K. Electrical conductivity through single molecules of XYL is measured for a Au/XYL/Au structure with chemisorbed Au-S coupling at both contacts. Fabricating nanometer separated metal electrodes will be extremely useful for nanotechnology to build efficient nano/bio inspired devices by successfully probing the nanoscale materials (molecules, DNA, and nm clusters) for electrical studies, which is still a challenging job for today's science and technology.

\section*{Acknowledgment}
% optional entry into table of contents (if used)
%\addcontentsline{toc}{section}{Acknowledgment}

The authors would like to thank Jaewon Choi, Quingling Hang, and C. Y. Fong of School of Electrical and Computer Engineering, Purdue University, for helping in FESEM facility. This work was supported in part by NASA, Department of Energy and NSF, USA.

 
%\hfill November 18, 2002

\newpage

\section*{Figure Captions}

\baselineskip=24pt

\noindent {\bf Figure 1.}~(a) Molecular formula of a MPTMS molecule. (b) A schematic diagram of MPTMS monolayer adhesion that forms Si-O-Si covalent bonds with the SiO$_2$ surface silanols and the Au-S bond with the evaporated Au thin layers. (c) Microscopic image of a 200$\AA$ thick lithographically defined Au wire with a notch at the middle and connected to 4000$\AA$ thick Au pad layers at both ends for electrical probing.

\vspace{0.5in}

\noindent {\bf Figure 2.}~(a) Current - voltage characteristics in a Au wire during EIBJ. Inset [A] shows the threshold current and voltage values at break point for some EIBJ devices, and [B] shows the I-V characteristics of an empty gap, solid line is the Fowler Northem fit to the experimental data (open circles). (b) Histogram of the conductances observed in 125 nano-gap EIBJ (Au/Empty-gap/Au) devices.

\vspace{0.5in}

\noindent {\bf Figure 3.}~FESEM image of an EIBJ that shows a $\sim$1nm gap created between two Au electrodes. Although the Au wires are initially 2$\mu$m wide (fig.a), gap of length $\sim$1nm and width $\sim$5nm is observed (fig.b) after the break junction. Image (b) is the zooming picture of the selected portion in image (a).

\vspace{0.5in}

\noindent {\bf Figure 4.}~Room temperature I-V characteristics of Au/XYL/Au (squares) and Au/Empty-gap/Au (circles) structures.

\newpage

\begin{figure}

\vspace{2.5in}

\centering
\includegraphics[width=5in]{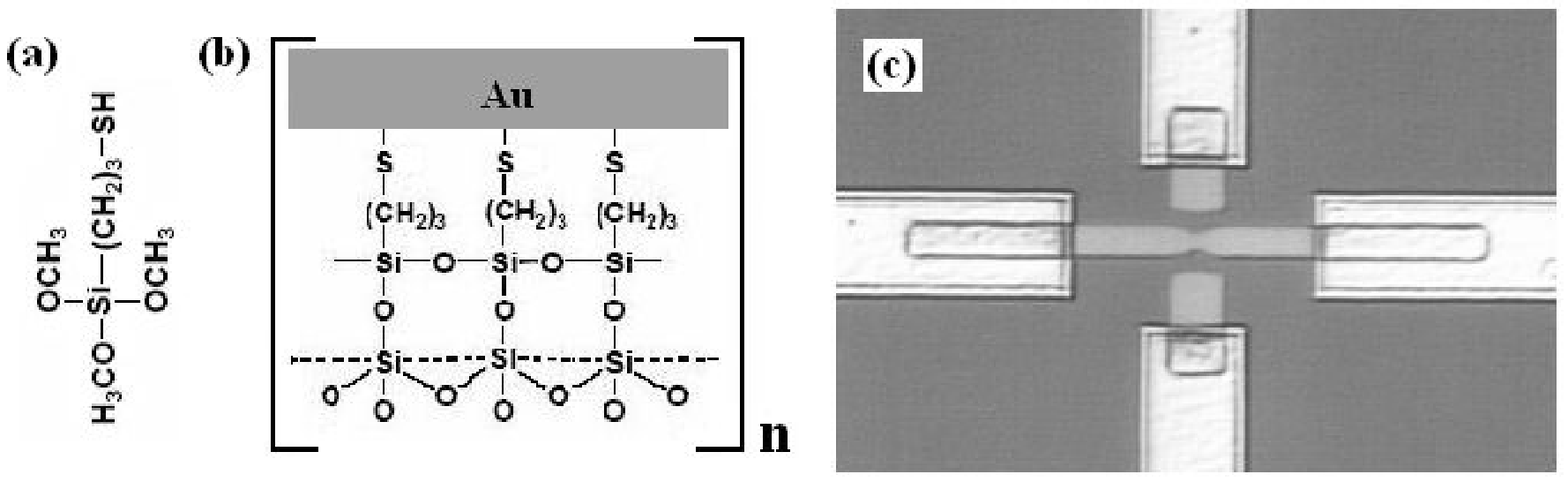}
%{expt_fig.eps}
%,height=1.1in
%\includegraphics[width=2.5in]{IEEE_nano_fig-1.eps}
\caption{}
%(a) Molecular formula of a MPTMS molecule. (b) A schematic diagram of MPTMS monolayer adhesion that forms Si-O-Si covalent bonds with the SiO$_2$ surface silanols and the Au-S bond with the evaporated Au thin layers. (c) Microscopic image of a 200$\AA$ thick lithographically defined Au wire with a notch at the middle and connected to 4000$\AA$ thick Au pad layers at both ends for electrical probing.}

\vspace{3.5in}

\label{figure-1}

\end{figure}

\newpage

\begin{figure}

\vspace{0.7in}

\centering
\includegraphics[width=5in]{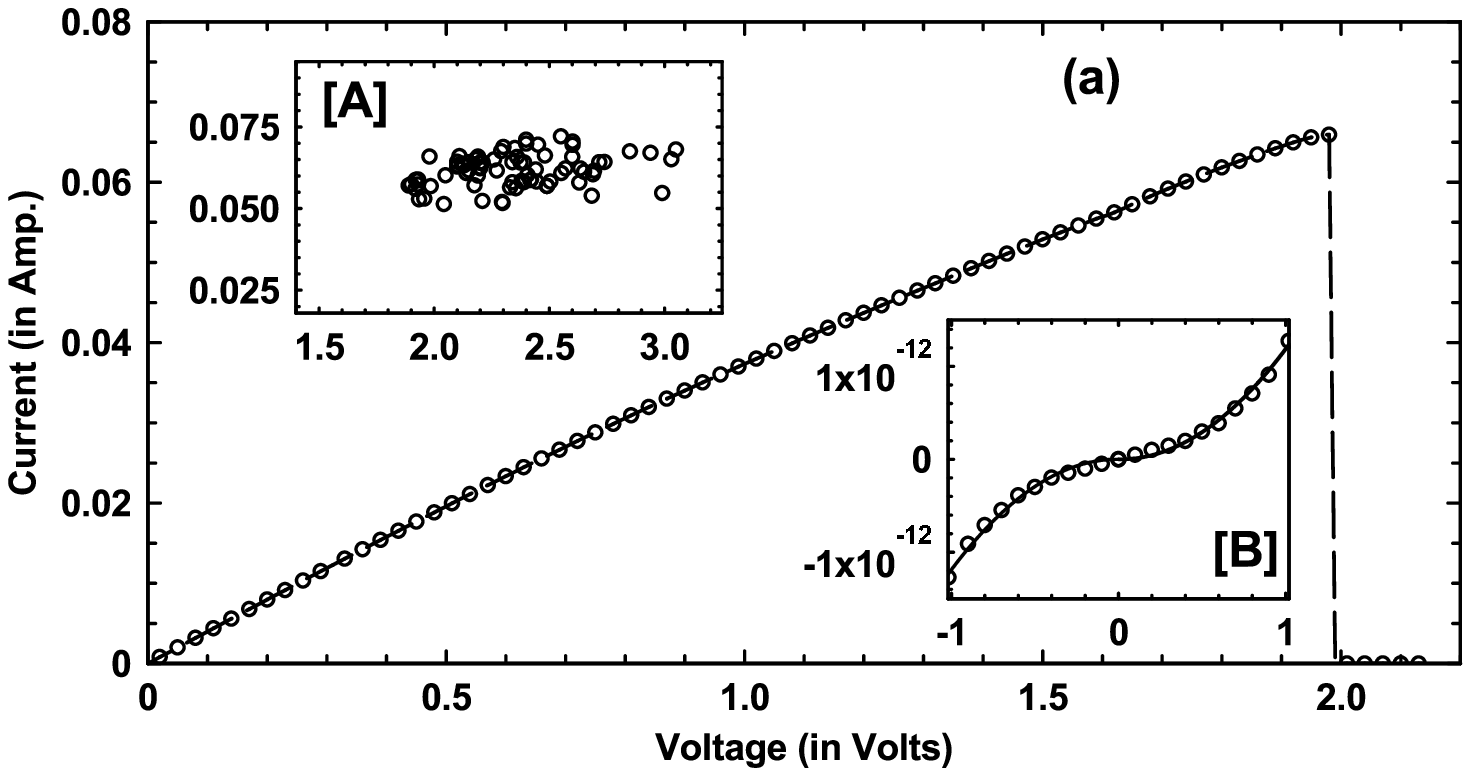}
%{fig-2a2.eps}
%{IEEE_nano_fig-2d.eps} 

\vspace{0.4cm}

\includegraphics[width=5in]{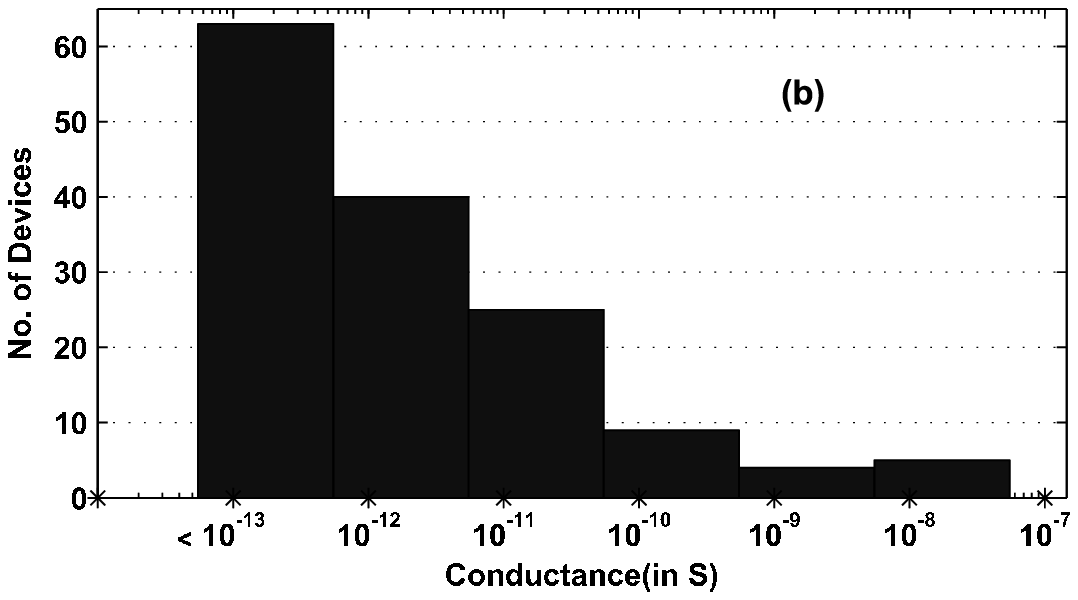}
%{tun_hist.eps}
\caption{}
%(a) Current - voltage characteristics in a Au wire during EIBJ. Inset [A] shows the threshold current and voltage values at break point for some EIBJ devices, and [B] shows the I-V characteristics of an empty gap, solid line is the Fowler Northem fit to the experimental data (open circles). (b) Histogram of the conductances observed in 125 nano-gap EIBJ (Au/Empty-gap/Au) devices.}

\vspace{1in}

\label{figure-2}
\end{figure}

\newpage

\begin{figure}

\vspace{1.5in}

\centering
\includegraphics[width=5in]{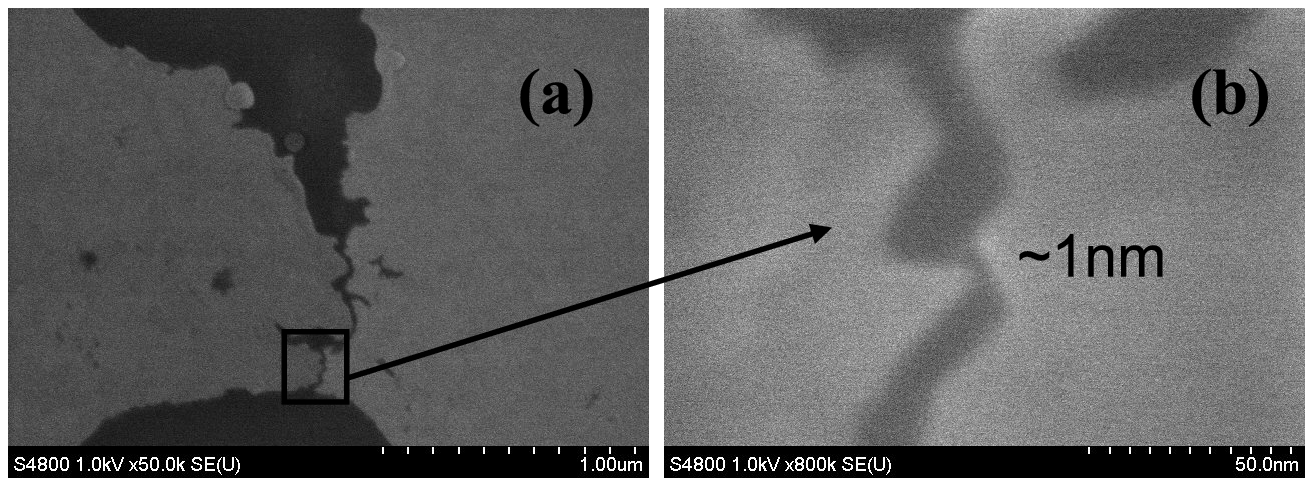}
%height=2.5in]{brk_FESEM_1.eps}
% where an .eps filename suffix will be assumed under latex, 
% and a .pdf suffix will be assumed for pdflatex
\caption{}
%FESEM image of an EIBJ that shows a $\sim$1nm gap created between two Au electrodes. Although the Au wires are initially 2$\mu m$ wide (fig.a), gap of length $\sim$1nm and width $\sim$5nm is observed (fig.b) after the break junction. Image (b) enlarged view of the selected portion in image (a).}

\vspace{2.2in}

\label{figure-3}
\end{figure}

\newpage

\begin{figure}
\centering
\includegraphics[width=5in]{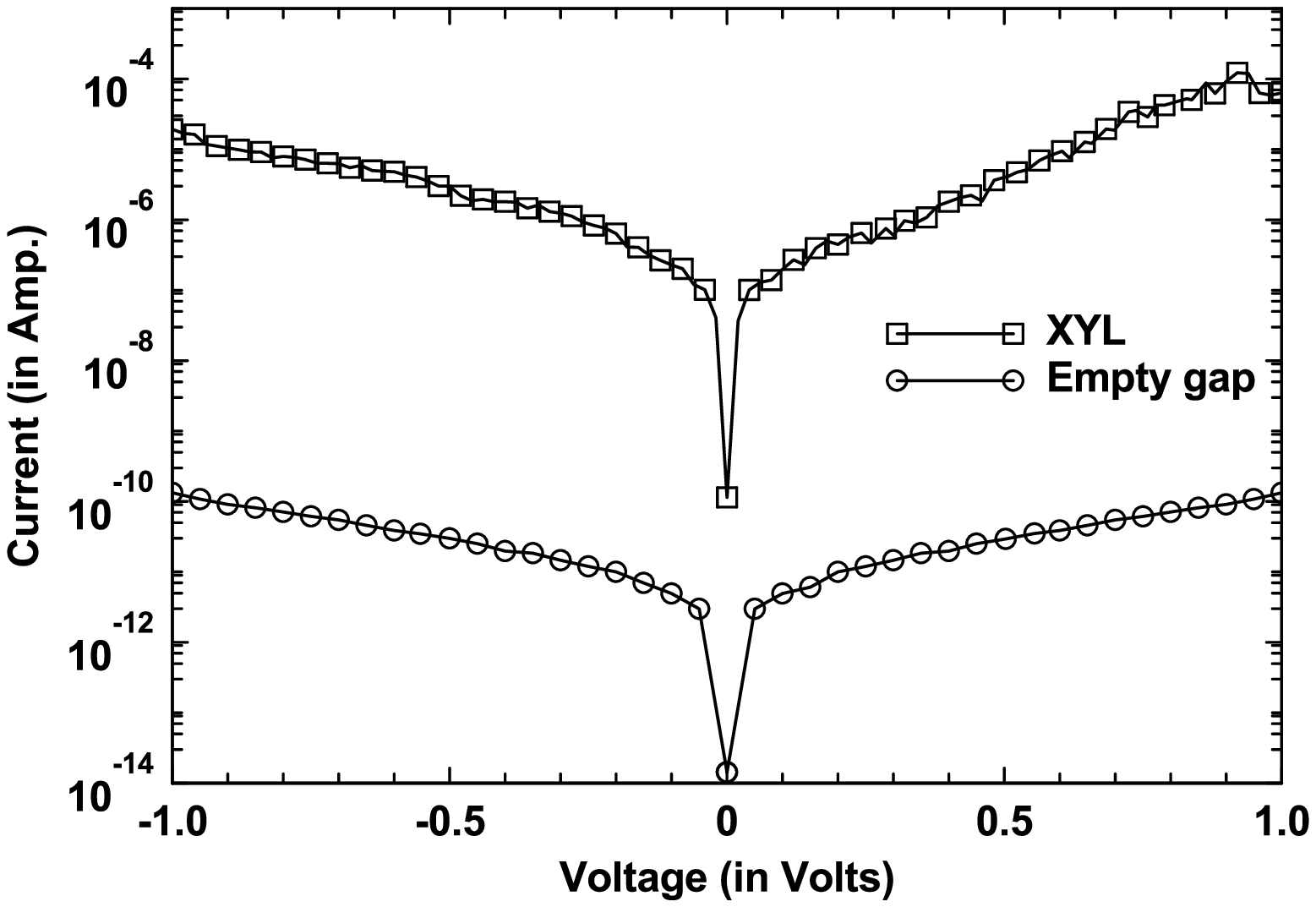}
\caption{ }
%Room temperature I-V characteristics of Au/XYL/Au (squares) and Au/Empty-gap/Au (circles) structures.}
\label{figure-4}
\end{figure}

\end{document}